\documentclass[runningheads]{llncs}
\usepackage[T1]{fontenc}
\usepackage{amsmath}
\usepackage{booktabs}
\usepackage{multirow}
\usepackage{amsmath}
\usepackage{graphicx}
\usepackage[table]{xcolor}
\usepackage{graphicx,verbatim}
\usepackage{amssymb}
\usepackage[pagebackref=true,breaklinks=true,letterpaper=true,colorlinks,citecolor=blue,linkcolor=blue,bookmarks=false]{hyperref}
\begin{document}
\title{PolypFlow: Reinforcing Polyp Segmentation with Flow-Driven Dynamics}
%

\author{Pu Wang\textsuperscript{1}, Huaizhi Ma\textsuperscript{2}, Zhihua Zhang\textsuperscript{1}, Zhuoran Zheng\textsuperscript{3}\thanks{Corresponding author: Zhuoran Zheng (zhengzr@njust.edu.cn)}}
\institute{
  $^1$\small Shandong University, School of mathematics, Jinnan 250100 , China \\
  $^2$\small  School of Artificial Intelligence Zaozhuang University, Zaozhuang 277100, China\\
  $^3$\small Sun Yat-sen University, School of cyber science and technology, 518000, China \\
  \email{202411943@mail.sdu.edu.cn, mahuaizhi@uzz.edu.cn, zhangzhihua@sdu.edu.cn, zhengzr@njust.edu.cn} }


\maketitle              
\begin{abstract}
Accurate polyp segmentation remains challenging due to irregular lesion morphologies, ambiguous boundaries, and heterogeneous imaging conditions. While U-Net variants excel at local feature fusion, they often lack explicit mechanisms to model the dynamic evolution of segmentation confidence under uncertainty. Inspired by the interpretable nature of flow-based models, we present \textbf{PolypFLow}, a flow-matching enhanced architecture that injects physics-inspired optimization dynamics into segmentation refinement. Unlike conventional cascaded networks, our framework solves an ordinary differential equation (ODE) to progressively align coarse initial predictions with ground truth masks through learned velocity fields. This trajectory-based refinement offers two key advantages: 1) Interpretable Optimization: Intermediate flow steps visualize how the model corrects under-segmented regions and sharpens boundaries at each ODE-solver iteration, demystifying the ``black-box" refinement process; 2) Boundary-Aware Robustness: The flow dynamics explicitly model gradient directions along polyp edges, enhancing resilience to low-contrast regions and motion artifacts. 
Numerous experimental results show that PolypFLow achieves a state-of-the-art  while maintaining consistent performance in different lighting scenarios.


\end{abstract}

\section{Introduction}
The precise segmentation of the polyp region is considered reliable measures to reduce CRC incidence and mortality~\cite{haggar2009colorectal,winawer1993prevention}. However, polyps exhibit high heterogeneity in size, color, texture, and boundary clarity, leading to a high rate of missed diagnoses~\cite{misawa2018artificial}. Developing high-precision polyp segmentation algorithms is of significant clinical importance for improving early CRC screening efficiency.

Currently, polyp segmentation algorithms can generally be classified into two categories. One class is low-parameter models based on convolutional~\cite{akbari2018polyp,ronneberger2015u} or Transformer networks~\cite{9845389}, they can be trained from scratch on a single 3090RTX GPU. These methods only use the inherent parameters for one or non-progressive continuous execution of inference, resulting in the problem of imprecise extraction of polyp contour information and blurred edge judgment.
Another approach is a polyp segmentation method based on the Segment Anything Model (SAM) for fine-tuning large models, such as Polyp-sam~\cite{li2024polyp} and Ployp-sam++~\cite{biswas2023polyp}. 
Since the pre-trained SAM has prior knowledge of the instance objects, it can be easily transferred to the polyp segmentation task and is more robust.
However, the method based on SAM has two drawbacks: 1) lack of interpretability, despite having a large number of instance level priors, it is difficult to understand its principle; 2) The number of parameters is large and struggle to deployed on edge devices. In addition, fine-tuning based on SAM also introduces uncertainty.

In order to solve the obstacles existing in the above problems, we propose the PolypFLow framework, which integrates U-Net with Flow Matching Equations (FME)~\cite{lipman2022flow}. Flow matching technology has prominent advantages in medical image analysis, its solution process can intuitively show how the model corrects under-segmented regions, featuring good interpretability. Further, PolypFLow introduces the discrete cosine transform (DCT) and the self-attention mechanism. DCT provides precise features for flow matching, and the self-attention mechanism focuses on global information, avoids local limitations, and significantly improves the segmentation accuracy. 
Our contributions are summarized as follows: 
\begin{enumerate}
    \item To the best of our knowledge, this paper is the first to introduce flow matching in the task of polyp image segmentation. It has good interpretability and effectively avoids the problems of inaccurate single inference of fixed-parameter models and the difficulty of deploying large-parameter-number models on edge devices.
    \item We develop a vector field that introduces self-attention and DCT, which can effectively capture the frequency domain and global information of the image. Extensive experiments on five benchmark datasets reveal that Polyp outperforms state-of-the-art methods.
\end{enumerate}

\section{Related Work}
Early polyp segmentation methods used feature-based machine learning models \cite{mamonov2014automated,tischendorf2010computer,zhou2021review}, extracted features from data and classified polyp using algorithms such as linear classifiers, k-Nearest Neighbors, and Support Vector Machines. 
However, these models can hardly capture the global context information and are not robust to complex scenarios. 
With the development of deep learning, several CNN-based polyp detection and segmentation models have been proposed \cite{tajbakhsh2015automated,zhang2018polyp}, these models have higher sensitivity compared to previous polyp detection models.
To locate the polyp boundaries with more precision, Fully Convolutional Networks (FCN) were applied to polyp segmentation \cite{akbari2018polyp,brandao2017fully}. 
U-Net is a classic medical image segmentation network, whose core architecture adopts a symmetric encoder-decoder structure and fuses low-level details with high-level semantic information through skip connections, achieving high-precision segmentation under limited data conditions \cite{ronneberger2015u}.
U-Net++ \cite{zhou2018unet++} and ResUNet++ \cite{jha2019resunet++} are improvements built upon U-Net and have been used for polyp segmentation, achieving good performance. 
However, these methods focus on segmenting the entire region of the polyp and neglect the boundary constraints of the region \cite{fan2020pranet}.

Flow Matching is a recent framework in generative modeling that has achieved state-of-the-art performance across various domains, including image, video, audio, and biological structures~\cite{lipman2024flow,yun2025flowhigh,lipman2022flow}. Weinzaepfel et al.~\cite{weinzaepfel2013deepflow} proposed a large-displacement optical flow computation method named DeepFlow, which addresses the performance bottlenecks of traditional approaches in scenarios with rapid motion and large displacements by integrating the deep matching algorithm into a variational optical flow framework. Shi et al.~\cite{shi2023flowformer++} introduced flow matching into the pre-training process for optical flow estimation, offering inspiration for the methodology presented in this work. Itai et al.~\cite{gat2025discrete} extended the applicability of flow matching by adapting its principles to discrete sequence data. Based on the advantages of the interpretability and efficient calculation of flow matching, we apply it to the polyp segmentation task. 

\begin{figure}[t]
    \centering
    \includegraphics[width=1\linewidth]{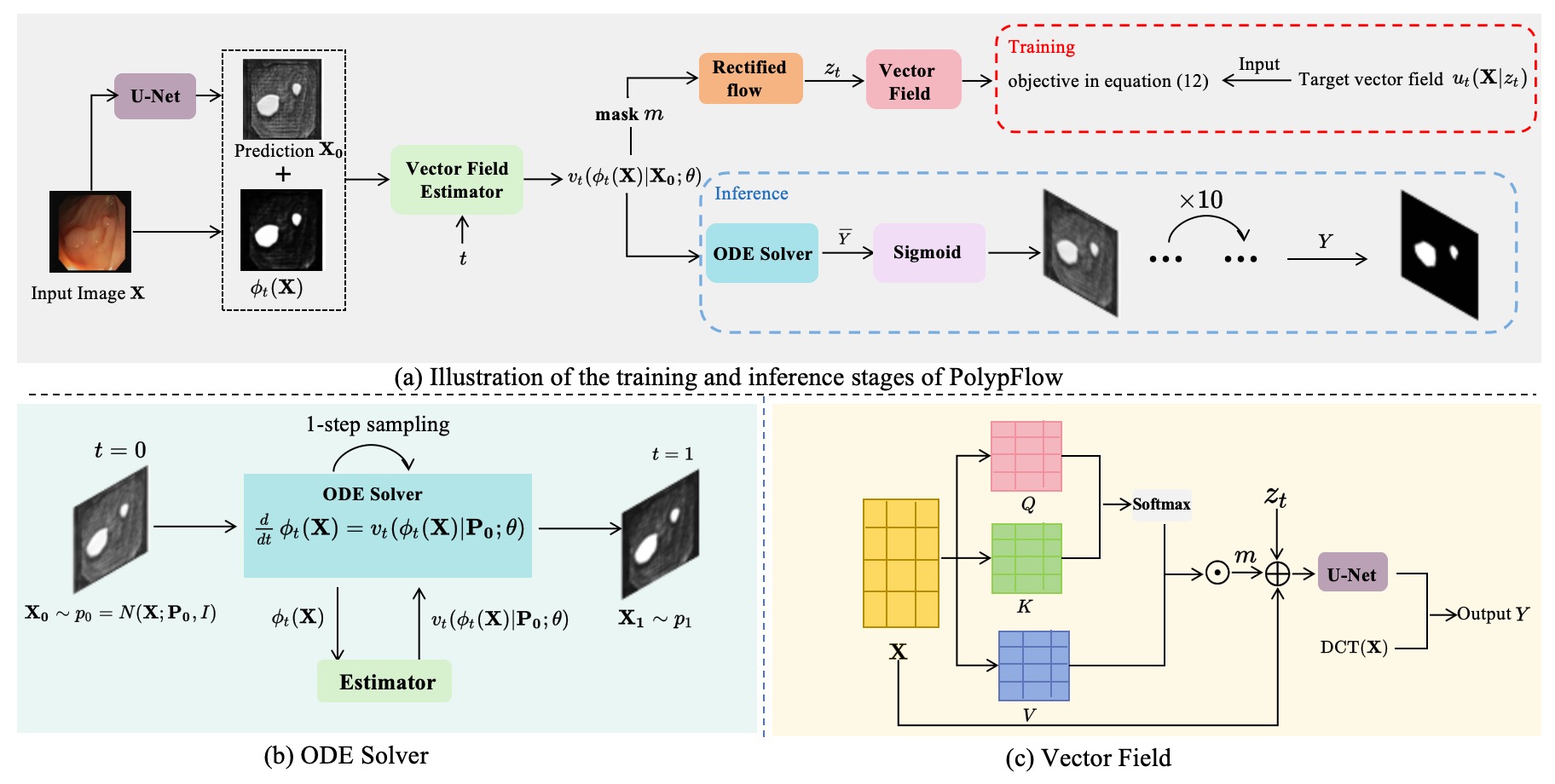}
    \caption{Overview of PolypFlow. (a) The overall training and inference process of PolypFlow based on flow matching. (b) The ordinary differential equation trajectory begins from a data-dependent prior distribution. (c) Vector field based on Self-attention and DCT. It is worth noting that our vector field is a core step in feature extraction, focusing on local features (convolution), global features (Self-attention), and the frequency domain (DCT).}
    \label{fig:model}
\end{figure}
\section{Method}
We construct PolypFLow by combining U-Net's  feature extraction capability with the  optimization dynamics of flow matching in Fig~\ref{fig:model}, more details are presented in the following subsections.
\subsection{Preliminaries}
Given the availability of empirical observations of the data distribution \(x_0\sim p_0\) and the noise distribution \(x_1\sim p_1\) (where \(p_1\) is typically Gaussian), the objective of the flow matching framework is to infer a coupling \(\pi(p_0,p_1)\) that characterizes the transformation between these two distributions. Continuous Normalizing Flows (CNFs)~\cite{chen2018neural} define a time-dependent probability density function, known as the probability density path \(p_t:[0,1]\times\mathbb{R}^d\rightarrow\mathbb{R}_{>0}\), with \(\int p_t(x)dx = 1\) and \(t\in[0,1]\). The flow, a time-dependent diffeomorphic mapping \(\phi_t:[0,1]\times\mathbb{R}^d\rightarrow\mathbb{R}^d\), which gives rise to the path \(p_t\), is propelled by a time-dependent vector field \(v_t:[0,1]\times\mathbb{R}^d\rightarrow\mathbb{R}^d\). This vector field is defined through the following ordinary differential equation (ODE):
\begin{equation}
   \frac{d}{dt}\phi_t(x)=v_t(\phi_t(x)),\quad\phi_0(x)=x\label{eq:ODE}
\end{equation}
Flow Matching seeks to optimize the simple regression objective:
\begin{equation}
  \mathcal{L}_{FM}(\theta) = \mathbb{E}_{t,p_t(x)}\left\|v_t(x;\theta)-u_t(x)\right\|^2\label{eq1}  
\end{equation}
In the above, $v_t(x;\theta)$ represents the parametric vector field for the CNFs, $u_t^\theta$ is the velocity with learnable parameter $\theta$, and \(u_t(x)\) is a vector field that generates a probability path \(p_t\). Euler's method is a commonly used ODE solution approach:
\begin{equation}
\mathbf{x}_{n + 1} = \mathbf{x}_n + v_t(\mathbf{x}_n)\Delta t\label{eq:Euler}
\end{equation}


\subsection{U-Net: Structure and Output}
U-Net and its variations have demonstrated remarkable efficiency in medical image segmentation tasks. In the encoder part, for a given image $\mathbf{I} \in \mathbb{R}^{H \times W \times 3}$, semantic information of images is gradually extracted through convolution to obtain a feature map:
\begin{equation}
\mathbf{E_1} = \mathrm{ReLU}(\mathrm{BN}(\mathrm{Conv}(\mathbf{I})))
\end{equation}
\begin{equation}
\mathbf{E_l} = \mathrm{ReLU}(\mathrm{BN}(\mathrm{Conv}(\mathbf{E_{l - 1}})))
\end{equation}
 where $\mathbf{E_l}$ is the $l-th \in \{1, 2, 3, 4\}$ layer encoder output, $\mathrm{ReLU}(\cdot)$ represents $\mathrm{ReLU}$ function, $\mathrm{BN}(\cdot)$ stands for batch normalization, $\mathrm{Conv}$ is a convolution. 
 In the decoder part, by transposing convolution and skipping join, encoder features $\mathbf{E_l}$ and upsampling features $\mathbf{D_{l + 1}}$ are spliced:
 \begin{equation}
    \mathbf{D_l} = \mathrm{ConvB}(\mathrm{Cat}(\mathrm{UpConv}(\mathbf{D_{l + 1}}), \mathbf{E_{4 - l}}))
 \end{equation}
 where $\mathrm{ConvB}(\cdot)$ denotes the convolution block and $\mathrm{Cat}(\cdot)$, $\mathrm{UpConv}(\cdot)$ are the feature concatenation operation and upsampling.
Here, we use a $1\times1$ convolution to get the model output.
So that low-level details and high-level semantic information are fused together to produce accurate segmentation results. 

\subsection{Flow Matching for Polyp Segmentation}
In the polyp segmentation task, to boost the model's ability to learn polyp features and segmentation accuracy, we propose a flow trajectories design method (vector field) that combines the mask and feature extractor.


\noindent\textbf{Design Vector Field.} The  vector field aims to capture the key features and relationships of polyps and their surrounding tissues:
\begin{equation}
\small
v_{\theta}(t, z_t, \mathbf{X}_{i}) = \underbrace{\text{Softmax}\left(\frac{QK^{T}}{\sqrt{d}}\right)V}_{\text{Self-attention weights}} \odot \left(\text{UNet}(z_t \oplus \mathbf{X}_{i} \oplus m) + \text{DCT}(\mathbf{X}_{i})\right)
\end{equation}
where $v_{\theta}$ is the vector field, representing the direction and degree of the model's attention to image features in different states. $t$ represents the time step, reflecting the stage changes of the model during training. $z_t$ is a latent variable that stores the information of the model's intermediate process. $\mathbf{X}$ is the input medical image, containing the visual information of polyps and surrounding tissues, $m$ is the mask of the polyp. 
$\text{DCT}(\mathbf{X})$ is the discrete cosine transform of the image $\mathbf{X}$, introducing frequency-domain prior knowledge to assist the model in exploring the potential structure of the image. $\odot$ represents element-wise multiplication. Through this operation, the self-attention weights are integrated with the feature information to generate the final vector field.
It is worth noting that the output of U-Net is encoded into $Q$, $K$, and $V$ by a large-scale convolutional kernel for self-attention, and $m$ is a learnable parameter matrix.

\begin{figure}[h]
    \centering
    \includegraphics[width=1\textwidth]{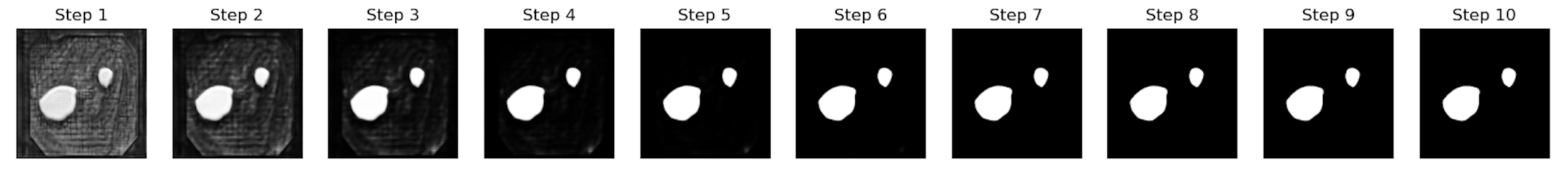}
    \caption{Visualize the results at each step.}
    \label{fig:step}
\end{figure}

After obtaining the vector field $v_\theta$, we use Eq.~\ref{eq:Euler} to update the input polyp images $\mathbf{X}$. 
Through multiple iterations, we can gradually transform the input polyp images $\mathbf{X_0}$ to $\mathbf{X_N}$ that is closer to the target distribution. The number of integration steps $N$ is set to 10, s shown in Figure~\ref{fig:step}. Our loss function can be formulated as:
\begin{equation}
    \mathcal{L} = \mathcal{L}_{IoU}^{w} + \mathcal{L}_{BCE}^{w}
\end{equation}
where $\mathcal{L}_{IoU}^{w}$ and $\mathcal{L}_{BCE}^{w}$ represent the weighted intersection over union loss and weighted binary cross entropy loss~\cite{zheng2024polyp}. The loss function enables the model to focus on boundary regions while optimizing the overlap between predictions and ground truth labels in the segmentation task. 

\section{Experiment}
\begin{table}[t]
\centering
\caption{Quantitative results on Kvasir and ClinicDB datasets.}
\label{table:1}
\fontsize{15pt}{7pt}\selectfont
\setlength{\tabcolsep}{8pt}
\resizebox{\textwidth}{!}{
\begin{tabular}{ll|c|cccccc}
\toprule
\multicolumn{2}{c|}{Methods} & year & mDice & mIoU & $F_\beta^\omega$ & $S_\alpha$ & $E_\phi^{\text{max}}$ & MAE \\
\midrule
\multirow{6}{*}{\rotatebox{90}{Kvasir-SEG}} & U-Net~\cite{ronneberger2015u} & 2015 & 0.818 & 0.746 & 0.794 & 0.858 & 0.893 & 0.055 \\
\multirow{6}{*}{} & U-Net++~\cite{zhou2018unet++} & 2018 & 0.821 & 0.744 & 0.808 & 0.862 & 0.909 & 0.048 \\
\multirow{6}{*}{} & PraNet~\cite{fan2020pranet} & 2020 & 0.898 & 0.841 & 0.885 & 0.915 & 0.948 & 0.030 \\
\multirow{6}{*}{} & SANet~\cite{wei2021shallow} & 2021 & 0.904 & 0.847 & 0.892 & 0.915 & 0.953 & 0.032 \\
\multirow{6}{*}{} & MedSAM~\cite{ma2024segment} & 2024 & 0.862 & 0.795 & - & - & - & - \\
\multirow{6}{*}{} & $\text{M}^2$ixNet~\cite{zheng2024polyp} & 2024 & 0.929 & 0.881 & 0.929 & 0.935 & 0.971 & 0.014 \\
\multirow{6}{*}{} & CAFE-Net~\cite{liu2024cafe} & 2024 & 0.933 & 0.889 & 0.927 & 0.939 & 0.971 & 0.019 \\
\multirow{6}{*}{} & \textbf{PolypFlow(Ours)} & 2025 & \textbf{0.971} & \textbf{0.948} & \textbf{0.967} & \textbf{0.964} & \textbf{0.993} & \textbf{0.008} \\
\midrule
\midrule
\multirow{6}{*}{\rotatebox{90}{ClinicDB}} & U-Net~\cite{ronneberger2015u} & 2015 & 0.823 & 0.755 & 0.811 & 0.890 & 0.953 & 0.019\\
\multirow{6}{*}{} & U-Net++~\cite{zhou2018unet++} & 2018 & 0.749 & 0.729 & 0.785 & 0.873 & 0.931 & 0.022\\
\multirow{6}{*}{} & PraNet~\cite{fan2020pranet} & 2020 & 0.899 & 0.849 & 0.896 & 0.937 & 0.979 & 0.009\\
\multirow{6}{*}{} & SANet~\cite{wei2021shallow} & 2021 & 0.916 & 0.859 & 0.909 & 0.940 & 0.976 & 0.012 \\
\multirow{6}{*}{} & MedSAM~\cite{ma2024segment} & 2024 & 0.867 & 0.803 & - & - & - & - \\
\multirow{6}{*}{} & $\text{M}^2$ixNet~\cite{zheng2024polyp} & 2024 & 0.941 & 0.891 & 0.934 & 0.955 & 0.986 & 0.005 \\
\multirow{6}{*}{} & CAFE-Net~\cite{liu2024cafe} & 2024 & 0.943 & 0.899 & \textbf{0.941} & 0.957 & 0.986 & 0.009 \\
\multirow{6}{*}{} & \textbf{PolypFlow(Ours)} & 2025 & \textbf{0.951} & \textbf{0.900} & 0.929 & \textbf{0.961} & \textbf{0.987} & \textbf{0.004} \\
\bottomrule
\end{tabular}
}
\end{table}
\subsection{Datasets and Training Settings.}
We compare the proposed PolypFlow with previous state-of-the-art methods, such as U-Net~\cite{ronneberger2015u}, U-Net++~\cite{zhou2018unet++}, PraNet~\cite{fan2020pranet}, SANet~\cite{wei2021shallow}, MedSAM~\cite{ma2024segment}, CAFE-Net~\cite{liu2024cafe}, and $\text{M}^2$ixNet~\cite{zheng2024polyp}. To assess the performance of the proposed PolypFLow. To this end, we adhere to the prevalent experimental setups in~\cite{zheng2024polyp}. Five publicly-accessible benchmarks are adopted: Kvasir-SEG~\cite{jha2020kvasir}, ClinicDB~\cite{bernal2015wm}, ColonDB~\cite{tajbakhsh2015automated}, Endoscene~\cite{vazquez2017benchmark}, and ETIS~\cite{silva2014toward}. In line with previous work~\cite{zheng2024polyp}, we utilize six widely-used metrics mDice, mIoU, MAE, $F_{\beta}^{w}$, $S_{\alpha}$, and $E_{\xi}^{max}$ for quantitative evaluation. 

The training set is composed of 900 images from Kvasir-SEG and 550 images from CVC-ClinicDB. 100 images from Kvasir-SEG and 62 images from CVC-ClinicDB are set aside as test sets. Since these images have been encountered during training, they are employed to evaluate the learning capacity of our model. We use ColonDB, Endoscene, and ETIS serve to measure the generalization ability of our model. Futher, PolypFlow is implemented using the PyTorch 2.0 framework. The GPU used is an NVIDIA TITAN RTX 4090 with 32G of RAM. During the training process, all input images are uniformly resized to 352×352. The network is trained for 100 epochs with a batch size of 8. For the optimizer, we choose AdamW to update the parameters. The learning rate is set to 1e-4. 

\begin{table}[h]
\centering
\caption{Quantitative results on ColonDB, Endoscene and ETIS datasets.}
\label{table:2}

\fontsize{15pt}{7pt}\selectfont
\setlength{\tabcolsep}{8pt}
\resizebox{\textwidth}{!}{
\begin{tabular}{ll|c|cccccc}
\toprule
\multicolumn{2}{c|}{Methods} & year & mDice & mIoU & $F_\beta^\omega$ & $S_\alpha$ & $E_\phi^{\text{max}}$ & MAE \\
\midrule
\multirow{6}{*}{\rotatebox{90}{ColonDB}} & U-Net~\cite{ronneberger2015u} & 2015 & 0.504 & 0.436 & 0.491 & 0.710 & 0.781 & 0.059 \\
\multirow{6}{*}{} & U-Net++~\cite{zhou2018unet++} & 2018 & 0.481 & 0.408 & 0.467 & 0.693 & 0.763 & 0.061 \\
\multirow{6}{*}{} & PraNet~\cite{fan2020pranet} & 2020 & 0.712 & 0.640 & 0.699 & 0.820 & 0.872 & 0.043 \\
\multirow{6}{*}{} & SANet~\cite{wei2021shallow} & 2021 & 0.752 & 0.669 & 0.725 & 0.837 & 0.875 & 0.043 \\
\multirow{6}{*}{} & MedSAM~\cite{ma2024segment} & 2024 & 0.734 & 0.651 & - & - & - & - \\
\multirow{6}{*}{} & $\text{M}^2$ixNet~\cite{zheng2024polyp} & 2024 & 0.820 & \textbf{0.855} & 0.799 & 0.869 & 0.935 & \textbf{0.021} \\
\multirow{6}{*}{} & CAFE-Net~\cite{liu2024cafe} & 2024 & 0.820 & 0.740 & 0.803 & 0.874 & 0.918 & 0.026 \\
\multirow{6}{*}{} & \textbf{PolypFlow(Ours)} & 2025 & \textbf{0.822} & 0.741 & \textbf{0.810} & \textbf{0.877} & \textbf{0.952} & 0.022 \\
\midrule
\midrule
\multirow{6}{*}{\rotatebox{90}{Endoscene}} & U-Net~\cite{ronneberger2015u} & 2015 & 0.710 & 0.627 & 0.684 & 0.843 & 0.875 & 0.022\\
\multirow{6}{*}{} & U-Net++~\cite{zhou2018unet++} & 2018 & 0.707 & 0.624 & 0.687 & 0.839 & 0.898 & 0.018\\
\multirow{6}{*}{} & PraNet~\cite{fan2020pranet} & 2020 & 0.871 & 0.797 & 0.843 & 0.925 & 0.972 & 0.010\\
\multirow{6}{*}{} & SANet~\cite{wei2021shallow} & 2021 & 0.888 & 0.815 & 0.859 & 0.928 & 0.972 & 0.008 \\
\multirow{6}{*}{} & MedSAM~\cite{ma2024segment} & 2024 & 0.870 & 0.798 & - & - & - & - \\
\multirow{6}{*}{} & $\text{M}^2$ixNet~\cite{zheng2024polyp} & 2024 & 0.895 & \textbf{0.861} & 0.879 & 0.941 & 0.978 & \textbf{0.005} \\
\multirow{6}{*}{} & CAFE-Net~\cite{liu2024cafe} & 2024 & \textbf{0.901} & 0.834 & 0.882 & 0.939 & 0.981 & 0.006 \\
\multirow{6}{*}{} & \textbf{PolypFlow(Ours)} & 2025 & \textbf{0.901} & 0.852 & \textbf{0.885} & \textbf{0.953} & \textbf{0.986} & \textbf{0.005} \\
\midrule
\midrule
\multirow{6}{*}{\rotatebox{90}{ETIS}} & U-Net~\cite{ronneberger2015u} & 2015 & 0.398 & 0.335 & 0.366 & 0.684 & 0.740 & 0.036\\
\multirow{6}{*}{} & U-Net++~\cite{zhou2018unet++} & 2018 & 0.401 & 0.343 & 0.390 & 0.683 & 0.776 & 0.035\\
\multirow{6}{*}{} & PraNet~\cite{fan2020pranet} & 2020 & 0.628 & 0.567 & 0.600 & 0.794 & 0.841 & 0.031\\
\multirow{6}{*}{} & SANet~\cite{wei2021shallow} & 2021 & 0.750 & 0.654 & 0.685 & 0.849 & 0.897 & 0.015 \\
\multirow{6}{*}{} & MedSAM~\cite{ma2024segment} & 2024 & 0.687 & 0.604 & - & - & - & - \\
\multirow{6}{*}{} & $\text{M}^2$ixNet~\cite{zheng2024polyp} & 2024 & 0.891 & 0.866 & 0.753 & 0.882 & 0.933 & \textbf{0.009} \\
\multirow{6}{*}{} & CAFE-Net~\cite{liu2024cafe} & 2024 & 0.822 & 0.738 & 0.775 & \textbf{0.898} & 0.940 & 0.014 \\
\multirow{6}{*}{} & \textbf{PolypFlow(Ours)} & 2025 & \textbf{0.895} & \textbf{0.872} & \textbf{0.815} & 0.895 & \textbf{0.972} & 0.010 \\

\bottomrule
\end{tabular}
}
\end{table}

\subsection{Performance Comparisons}
we conduct two experiments to validate our model’s learning ability on two seen datasets, i.e., Kvasir and ClinicDB. As shown in Tab.~\ref{table:1}, PolypFlow outperforms all baselines by a large margin (mDice > 4\%) across both two datasets, in all metrics. This suggests that our model has a strong learning ability to effectively segment polyps.

\begin{figure}[h]
    \centering
    \includegraphics[width=1\linewidth]{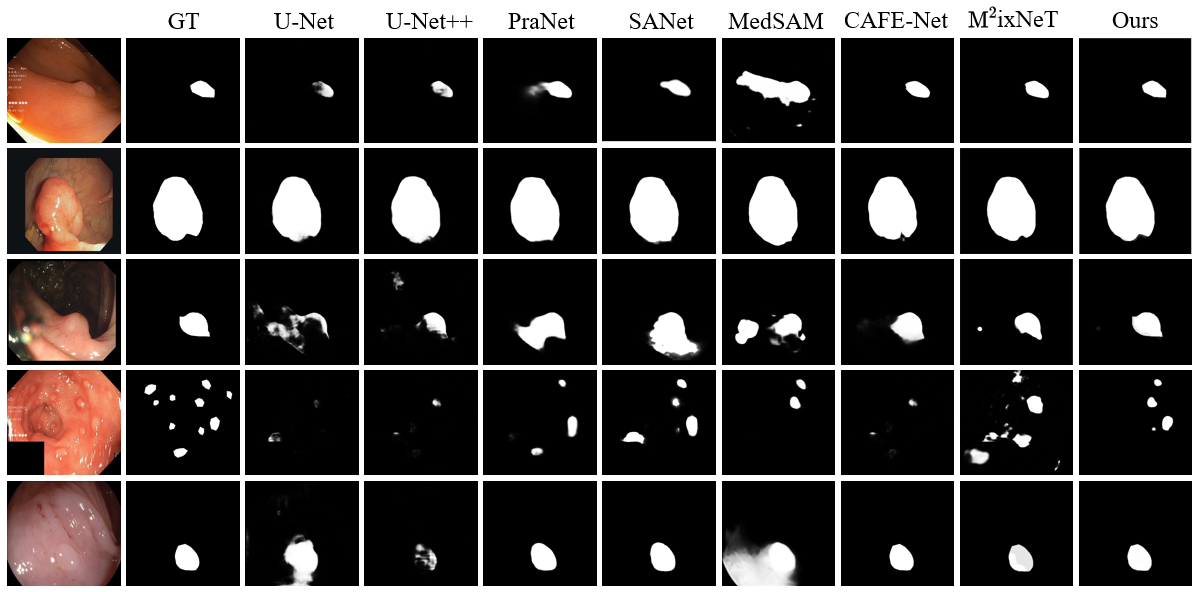}
    
    \caption{visual results of different methods.}
    \label{fig:vision}
\end{figure}

We conduct three experiments to test the model’s generalizability. PolypFlow outperforms existing classical medical segmentation baselines (U-Net, U-Net++), with significant improvements on all three unseen datasets in Tab.~\ref{table:2}.
Fig.~\ref{fig:vision} presents the visualization results of our model and the compared models. Our proposed model outperforms other methods significantly, with segmentation results closest to the ground truth. 
\begin{table}[h]
\centering
\caption{Ablation study.}
\label{table:3}

\fontsize{8pt}{6.5pt}\selectfont
\setlength{\tabcolsep}{8pt}
\resizebox{\textwidth}{!}{
\begin{tabular}{r|cc|cc}
\toprule
Settings & \multicolumn{2}{c|}{ClinicDB (seen)} & \multicolumn{2}{c}{Endoscene (unseen)}\\

& mDice & mIoU
& mDice & mIoU
\\
\midrule 
Backbone & 0.736 & 0.568 
& 0.721 & 0.607
\\
SA + Backbone & 0.835 & 0.743 
& 0.863 & 0.786
\\
DCT($\cdot$)+ Backbone & 0.875 & 0.847
&0.881 & \textbf{0.859}
\\
SA +DCT($\cdot$) + Backbone & \textbf{0.901} & \textbf{0.885} 
& \textbf{0.891} & 0.797
\\

\bottomrule
\end{tabular}
}
\end{table}
\subsection{Ablation Study and Discussion}
We conduct tests on every component of our PolypFlow that the model has encountered during seen datasets and unseen datasets to offer a more profound understanding of our model's performance and characteristics. 
The experimental results demonstrate that the self-attention module (SA) significantly improves model performance. From Tab.~\ref{table:3}, on the ClinicDB (seen dataset), after adding SA, the mDice and mIoU increase from 0.736 and 0.568 to 0.835 and 0.743, respectively. On the Endosceme (unseen dataset), the mDice and mIoU increase from 0.721 and 0.607 to 0.863 and 0.786, respectively. This indicates that SA effectively captures global contextual information, thereby enhancing segmentation accuracy.
The DCT($\cdot$) also plays a crucial role in improving model performance.  From Tab.~\ref{table:3}, on the ClinicDB (seen dataset), after adding DCT, the mDice and mIoU increase to 0.875 and 0.847, respectively. On the Endosceme (unseen dataset), the mDice and mIoU increase to 0.881 and 0.859, respectively. By extracting frequency-domain features, DCT enhances the model's ability to capture detailed information, thereby further optimizing segmentation results.

Here, we have conducted two additional experiments to verify the effectiveness of our method. 1: We selected iterative vector fields with 1, 5, 8, and 15 steps. We found that increasing the number of steps can improve the model's performance, but the difference between 15 steps and 10 steps is minimal. In addition, we replaced U-Net with U-Net++, which improved the performance of polyp segmentation by 2.3\%.

\section{Conclusion}
In this work, we develop PolypFlow, for automatically segmenting polyps from colonoscopy images, which a vector field based on self-attention and DCT. Extensive experimental evaluations have shown that, on five challenging datasets, PolypFlow consistently surpasses all existing state-of-the-art methods with a significant margin, achieving an improvement of more than 4.5\%. Another advantage is that PolypFlow visualizes the process of polyp segmentation.

\bibliographystyle{splncs04}
\bibliography{ref}

\end{document}